\documentstyle[seceq,epsf,wrapft]{ptptex}
\def\lsim{\mathrel{\mathpalette\Oversim<}}
\def\gsim{\mathrel{\mathpalette\Oversim>}}
\def\Oversim#1#2{\lower0.5ex\vbox{\baselineskip0pt\lineskip0pt%
            \lineskiplimit0pt\ialign{%
          $\mathsurround0pt #1\hfil##\hfil$\crcr#2\crcr\sim\crcr}}}

\def\Hbun{H$_2$~}
\def\H2p{H$_2^+$~}

\def\Hm{H$^-$}
\def\sun{\odot}
\def\be{\begin{equation}}
\def\ee{\end{equation}}
\def\ba{\begin{eqnarray}}
\def\ea{\end{eqnarray}}

\title{Thermal and Dynamical Evolution of Primordial Gas Clouds}
\subtitle{On the Formation of First Luminous Objects}    
\author{
Ryoichi {\sc Nishi},\footnote{E-mail: nishi@tap.scphys.kyoto-u.ac.jp} 
Hajime {\sc Susa},$^{*,}$\footnote{E-mail: susa@rccp.tsukuba.ac.jp}
Hideya {\sc Uehara},$^{**,}$\footnote{E-mail: uehara@th.nao.ac.jp}\\
Masako {\sc Yamada},\footnote{E-mail: masako@tap.scphys.kyoto-u.ac.jp}
and Kazuyuki {\sc Omukai}\footnote{E-mail: omukai@tap.scphys.kyoto-u.ac.jp}
}

\inst{%         %Affiliation, neglected when [addenda] or [errata]
Department of Physics, Kyoto University, Kyoto 606-8502, Japan\\
$^*$Center for Computational Physics, University of Tsukuba\\
Tsukuba 305-8571, Japan\\
$^{**}$Division of Theoretical Astrophysics, 
National Astronomical Observatory\\
Mitaka 181-8588, Japan
}

\recdate{%      %Editorial Office will fill in this.
September 21, 1998
}

\abst{%       %this abstract is neglected when [addenda] or [errata]
We investigate the thermal and dynamical evolution of 
primordial gas clouds in the universe after decoupling.
Comparing the time-scale of dynamical evolution with 
that of fragmentation,  
we can estimate the typical fragmentation scale. 
We propose the following scenario of the formation process of 
first luminous objects consisting of large number stars. 
First, by pancake collapse of the overdensity regions 
in the expanding 
universe or collision between clouds in potential wells, 
quasi-plane shocks form. 
If the shock-heated temperature is higher than about $10^4$ K, 
the postshock gas cools down to several hundred K by \Hbun line cooling, 
and the shock-compressed layer fragments into filamentary clouds. 
The filamentary cloud collapses dynamically once more 
and fragments into cloud cores. 
Finally, a primordial star forms in a cloud core. 
We show that  
the minimum mass of the first star is essentially determined 
by the {\it Chandrasekhar mass}.  
Also, we investigate the dynamical collapse of cloud cores by numerical 
simulation and show that the evolution paths of the central regions 
of the cores depend only very weakly on the total core mass. 
After mass accretion, a massive star may be formed in a core, 
since the estimated mass accretion rate is very large. 
In such a case, it may be possible for many massive stars 
form almost simultaneously. Then the clouds can be luminous objects.
On the other hand, if the shock-heated temperature is lower, 
effective star formation is delayed significantly. 
}

\begin{document}

\maketitle
\section{Introduction}

Today, we have a great deal of observational data concerning the 
early universe. 
However, we have very little information about an era referred to as 
the `dark ages'.
Information regarding a era of recombination 
(with redshift $z$ of about $10^3$) can be obtained 
by observation of cosmic microwave background radiation. 
After recombination we can obtain little information until $z \sim 5$, 
when we can observe luminous objects such as galaxies and QSOs. 
On the other hand, formation stage of these objects should be 
in the `dark ages', and thus investigating the formation process 
of galaxies using a theoretical approach is very important. 

In the standard theory of the formation of luminous objects, 
the most basic idea is that they are made from giant gas
clouds, which were once slightly overdense regions in the early epoch  
and grew to depart from the general cosmic expansion. 
After contraction of these clouds, if they fragment into many 
stellar-size clouds and many massive stars are formed 
on a time scale of the life 
of the massive stars ($10^7$ yr), they can become luminous objects. 
To understand the way in which luminous objects are formed, we 
investigate physical processes of the clouds in various stages of 
evolution.

To discuss the formation and evolution of galaxies,  
it is necessary to investigate star formation processes in
protogalactic clouds, since stars are very important components of 
observed galaxies.
In addition, stars play crucial roles in galactic activities, 
for example, ultraviolet radiation and supernova explosions, which 
are important in some cosmological contexts.
Moreover, considering the hierarchical clustering scenario, 
several authors have stressed the importance of stars 
in pregalactic objects  
for the reionization of the universe (e.g., Ref.~\citen{rf:1}). 
Therefore, to study galaxy formation and evolution of 
intergalactic matter, it is necessary to investigate 
the star formation process in the early universe. 

Because gravity alone cannot be used to derive natural scales, 
to form objects of various scales, 
such as galaxies and stars, thermal processes
within gases are very important.
In the early epoch, before the first stars were formed, 
baryonic matter was made mostly of hydrogen and helium, 
and the most important coolants
at relatively low temperature was hydrogen molecules. 
Without hydrogen molecules, gas clouds of primordial composition are 
ineffective to cool below about $10^4$~K, because of the high energy of
hydrogen Ly$\alpha$ photons. 
Hence the Jeans mass of a purely atomic gas of primordial composition is
about $M_J\propto T^{3/2}\rho^{-1/2}\sim 10^7(T/10^4~{\rm K})^{3/2}
(n/1~{\rm cm}^{-3})^{-1/2}M_{\sun}$, far larger than the stellar scale. 
Sufficiently after the recombination ($z_{\rm rec}\sim 1100$), the
fraction of H$_2$ is frozen at a constant value of order 
$10^{-5}$,\cite{GP} and 
this small amount of H$_2$ is not enough to cool the gas in order to lower 
the Jeans mass down to stellar mass within the age of the universe.
Hence the rength scale into which the clouds fragment depends strongly 
on the amount of H$_2$ which is formed in the 
clouds during their contraction.

Because of the absence of dust grains and the inefficiency of direct
reaction (H + H $\to$ H$_2$ +$\gamma$), H$_2$ molecules are formed 
through the two indirect paths described below: 

\ba
{\rm H}+ e &\to& {\rm H}^- +\gamma \label{H-},\\
{\rm H}+{\rm H^-} &\to& {\rm H}_2 +e, 
\ea

\ba
{\rm H}+{\rm H}^+ &\to& {\rm H}_2^+ +\gamma, \\
{\rm H}_2^+ +{\rm H} &\to& {\rm H}_2 + {\rm H}^+.
\label{H2+}
\ea
Both paths require a certain degree of ionization, and usually the 
first path is more effective than second path. 
Hydrogen molecules 
are collisionally dissociated efficiently at $T \gsim$ several 
thousand K. 
Then, in order for H$_2$ to be made through the above two reactions, 
there must be some degree of ionization at $T \lsim 10^4$ K.
On the other hand, if in chemical equilibrium, there remain
few electrons or H$^+$ ions at $T \lsim 10^4$ K. 
In such clouds, little \Hbun can be formed. 
When $z \gsim 10^2$, H$^-$ is radiatively detached by cosmic background
radiation and \Hbun formation is inefficient.\cite{GP,MST} 
Thus, we consider clouds collapsing only when $z \lsim 10^2$.  

In addition, the thermal evolution of 
clouds is strongly affected by the initial shape of the clouds,
or their angular momentum in the initial state.\cite{Hut}
The collapse of an initially oblate cloud is very different 
from that of spherical one. 
The non-spherical collapse in the initial phase is immediately 
followed by a disk-like collapse, which has a different time scale
from a spherical collapse. 
Strictly speaking, the times needed for $\rho$ to be $\infty$ 
do not differ greatly from each other,
but {\it at the same density}, which is much higher than the initial value, 
the time scales determined 
by the temporal shrinking rates are entirely different.  
For spherical collapse we have 
\be
t_{\rm dyn} \equiv -r/v_r =    \propto \rho^{-1/2}
~({\rm free ~fall ~time}), \label{eqn:sphere}
\ee
and for disk-like collapse, 
\be
t_{\rm dyn} \equiv \; -z/v_z   \propto \rho^{-1} ~(z \rightarrow 0). 
\label{eqn:sheet} 
\ee
Since, after sufficient collapse, $v_z$ is almost constant 
and $z$ changes as $\rho^{-1}$.
Then $t_{\rm dyn}$ for disk-like collapse becomes much shorter 
than the free-fall time at the same density.
And this time scale $t_{\rm dyn}$ is also the time scale of 
adiabatic heating.
So the thermal evolution of these oblate spheroids is also 
very different from that of spherical clouds.
Moreover, thin disk-like clouds can fragment into smaller clouds. 
Then it is necessary to include the effect of shape change 
when we investigate the thermal evolution and the fragmentation 
of clouds. 

In this paper, we put together important physical processes regarding 
cloud evolution in the early universe in various evolutional stages.  
In \S 2 we investigate the thermal and dynamical evolution of primordial 
gas clouds and discuss fragmentation processes. 
In \S 3 we disscuss cylindrical collapse, which follows fragmentation 
of the disk-like cloud. 
In \S 4 we investigate the evolution of cloud cores 
and discuss the formation of stellar cores. 
Finally, we summarize our results 
and discuss the formation of first luminous objects in \S 5. 

\section{Contraction and fragmentation processes 
of primordial gas clouds}

In this section, we investigate the processes of contraction and 
fragmentation of primordial gas clouds. To discuss cloud evolution, 
we compare various time scales. 
And we consider that 
a collapsing cloud fragments when the condition
\begin{equation}
t_{\rm dyn} \sim t_{\rm frag}    ,     \label{s1}
\end{equation}
is satisfied, where
\begin{eqnarray*}
& t_{\rm dyn}& \left ( \equiv \rho/ \frac{d \rho }{dt}\right ): 
\mbox{the timescale for dynamical evolution}, \\
& t_{\rm frag}& : \mbox{the timescale for fragmentation}.
\end{eqnarray*}

\subsection{Initial pressure-free collapse}

First, a cloud with a mass much larger than its own virial mass 
collapses almost as a free-fall collapse. 
Several authors investigated the thermal evolution of primordial clouds 
in pressure-free collapse with a 1-zone 
approximation.\cite{MST,Carl,PSS}
They also estimated the minimum Jeans mass of a collapsing cloud. 

However, spherical clouds in pressure-free collapse are 
unstable against non-spherical perturbations\cite{LMS} 
and eventually they collapse into a disk-like configuration.\cite{Hut} 
The effective `adiabatic index', $\Gamma \equiv 
{\partial \ln p / \partial t \over \partial \ln \rho / \partial t}$, 
changes with the collapse configuration, 
and, moreover, the critical $\Gamma$, which corresponds to 
occurence of bounce, also changes: 
it is 0 for disk collapse, 1 for cylindrical collapse, 
and 3/4 for spherical collapse.  
Thus we investigated the thermal evolution of non-spherical primordial 
gas clouds.\cite{SUN} 
We studied the infulence of the angular momentum and/or initial shape 
of the clouds in pressure-free collapse using a calculation of thermal 
evolution including radiative tansfer. 
We show that ordinary clouds bounce and are shock-heated 
when the hydrogen number density is much lower than $10^8$ cm$^{-3}$, 
which is the density at which three-body 
reaction for \Hbun formation becomes important. 
Therefore the Jeans mass cannot be lowered to  
the stellar mass scale in this stage.  

\subsection{Evolution of post schock gas}

As discussed above, ordinary pregalactic clouds collapse into disk-like 
configurations and are shock-heated when they bounce.  
Also, when subgalactic clouds collide, they are 
shock-heated. 
After considerable cooling, a shock-compressed layer 
can fragment into smaller clouds. 
Thus, we study the thermal evolution of the post-shock flow.\cite{Set}  

\subsubsection{Important time scales}

First, we compare the time scales which are essential to
the thermal evolution of the postshock layer in primordial gas
clouds. 
The various relevant time scales are 
\begin{eqnarray}
t_{\rm cool} & \equiv & C_p\frac{\rho k T}{\mu m_p \Lambda},
 \label{eqn:tcool}\\
t_{\rm ion}  & \equiv & \frac{n_p}{k^{\rm ion} n_{\rm H} n_e} 
= \frac{1}{k^{\rm ion} n_N (1-y_e)}, \label{eqn:tion}\\
t_{\rm rec}  & \equiv & \frac{n_p}{k^{\rm rec} n_e n_p} 
= \frac{1}{k^{\rm rec} n_N y_e}, \label{eqn:trec}\\
t_{\rm dis} & \equiv & \frac{n_{\rm H_2}}
{\dot n_{\rm H_2}^{\rm dis}}, \\
t_{\rm for} & \equiv & \frac{n_{\rm H_2}}
{\dot n_{\rm H_2}^{\rm for}}, 
\end{eqnarray}
where $t_{\rm cool},~t_{\rm ion},~t_{\rm rec},~t_{\rm dis}$ and 
$t_{\rm for}$ denote 
the time scales of cooling, ionization, recombination, ${\rm H_2}$
dissociation and formation, respectively.
\footnote{The time scale of thermalization in the post-shock layer is
shorter than the other time scales. For example, at $2\times 10^4~{\rm K}$
the time scales of thermalization for plasma and neutral gas are about
$10^2$ and $10^5$ times shorter than the ionization time scale, 
respectively.} 
$n_N$,~$n_i$,~$y_i ~(\equiv n_i / n_N)$,~$k^X$,~$\rho$,~$\mu$ 
and $m_p$ are the number density of nucleons,
the number density of the $i$th species, 
the fraction of $i$th species,
the chemical reaction rate coefficient of the ``$X$'' process,
the mass density, mean molecular weight and proton mass, respectively.
$\Lambda$ denotes the net cooling rate, which includes radiative and 
chemical cooling/heating. We assume that the system is 
optically thin for cooling photons. This assumption is valid for 
usual subgalactic clouds. 
For simplicity, we do not include helium. 
The existence of helium affects the mean molecular weight and the
cooling rate at high temperature $(T \sim 10^5~{\rm K})$. However,
these are minor effects on the thermal evolution of primordial gas
clouds at lower temperature ($T \lsim 10^4$ K), 
 which we are especially interested in. 
$C_p$ denotes the heat capacity of the gas cloud. 
Since we are interested in a quasi-steady post-shock flow, 
where $t_{\rm cool}$ is shorter than the time scale of the change of 
the shock velocity and/or pre-shock density, 
the post-shock layer is almost isobaric.
$\dot n_{\rm H_2}^{\rm dis}$ and $\dot n_{\rm H_2}^{\rm for}$ denote 
the effective \Hbun dissociation and formation rate. 
Since \Hbun molecules are formed through \Hm or \H2p 
(Eqs. (\ref{H-}) to (\ref{H2+})) 
and dissociated throgh ${\rm H_2^+}$,\cite{GP} we include only the part 
of reaction rate, that corresponds to the reaction 
formed from H and dissociated to H. 
Comparing these five time scales we can find the fastest
process without solving detailed time-dependent differential equations.

We should remark that {\it all of these time 
scales are proportional to $\rho^{-1}$} 
if the cooling rate and all reaction rates are proportional to 
$\rho^2$. 
The dominant component of the cooling changes according to 
temperature.
Below $10^4~{\rm K}$, ${\rm H_2}$ line emission dominates the total
cooling of the cloud. 
In this process, the cooling rate is proportional to the number density
of the hydrogen molecules at the excited levels whose fraction is 
proportional to $\rho$. 
\footnote{This treatment is limited to low density 
($n\lsim 10^4~{\rm cm^{-3}}$) 
The cooling rate is not proportional to $\rho^2$ for higher density 
(higher than the {\it critical density}).}.
Hence, the cooling time scale is proportional to $\rho^{-1}$ for 
$T \lsim 10^4~{\rm K}$, as in equation ($\ref{eqn:tcool}$). 
The cooling process in the temperature range 
$10^4~{\rm K} \lsim T \lsim 10^5~{\rm K}$
is dominated by the bound-bound transition of the hydrogen atoms,
and the number of excited hydrogen atoms is determined by 
collisions with other atoms. 
Therefore, the cooling rate is proportional to the square 
of the total density.
For higher temperatures ($T\gsim 10^5~{\rm K}$), the free-free emission 
by the collision between ions and electrons dominates the energy radiated 
away from the cloud.
In any case, the cooling rate is proportional to $\rho^2$, 
and the cooling time is proportional to $\rho^{-1}$. 
\footnote{Compton cooling is effective at high redshift and high temperature. 
In this case, the cooling rate is not proportional to $\rho^2$, 
but this does not change the following results significantly.}
The chemical reaction rates are proportional to $\rho^2$ if the reactions
are dominated by collisional processes. 

Photoionization and H$_2$
photodissociation processes caused by the UV photons emitted 
from the post-shock
hot region could affect the ionization rates and the H$_2$ dissociation
rates in post-shock regions. 
These radiative reactions are proportional to $\rho^2$, 
because the emission rate of UV photons in the post-shock
is proportional to $\rho^2$. 
We investigate 
the effects of UV photons from postshock regions.\cite{UN}  
However, as a 0th-order approximation, we 
neglect the photoionization and the H$_2$ photodissociation 
processes in this paper.

%%%%%%%%%%%%%%%%%%%%%%%%%%%%%%%%%%
% THE FRACTION OF H_2
%%%%%%%%%%%%%%%%%%%%%%%%%%%%%%%%%%
\subsubsection{Fraction of ${\rm H_2}$}

Here we consider the estimation of $y_{\rm H_2}$ in the post-shock flow 
for the case in which the post-shock temperature is higher than 
about $10^4$ K.
If $y_{\rm H_2}$ can be determined for given $T$ and $y_e$, 
the cooling time is always determined by just $T$ and $y_e$. 
As a result, the thermal evolution of the system is completely
determined in the $y_e$-$T$ plane. 

Now, we make some remarks regarding important properties 
of $t_{\rm dis}$ and $t_{\rm for}$. 
First, the time scale of dissociation is independent of $y_{\rm H_2}$, 
the fraction of hydrogen molecules.
On the other hand, $t_{\rm for}$ is proportional to $y_{\rm H_2}$.
Then, the time scales given in terms of their equilibrium
values are
\begin{eqnarray}
t_{\rm dis}& = &t_{\rm dis}^{\rm eq}, \label{eqn:diseq}\\
t_{\rm for}& = &t_{\rm for}^{\rm eq} 
\frac{y_{\rm H_2}}{y_{\rm H_2}^{\rm eq} (y_e, T)} 
=t_{\rm dis}^{\rm eq} \frac{y_{\rm H_2}}{y_{\rm H_2}^{\rm eq} (y_e, T)},
\label{eqn:foreq}
\end{eqnarray}
where the suffix eq denotes the value at which the hydrogen molecules
are in chemical equilibrium for a given electron abundance and
temperature. 
Thus, $t_{\rm dis} < t_{\rm rec},~t_{\rm cool}$ is a sufficient condition
$y_{\rm H_2}$ to be the equilibrium value for given $y_e$ and $T$.

%%%%%%%%%%%%%%%%%%%%%%%%%%%%%%%%%%%%%%%%%%%%%%%
%
% H_2 EQUILIBRIUM FOR RECOMBINATION
%
%%%%%%%%%%%%%%%%%%%%%%%%%%%%%%%%%%%%%%%%%%%%%%%

%%%%%% Figure2 %%%%%%%%%%%%%%
\begin{figure}
   \epsfxsize=8cm
   \centerline{\epsfbox{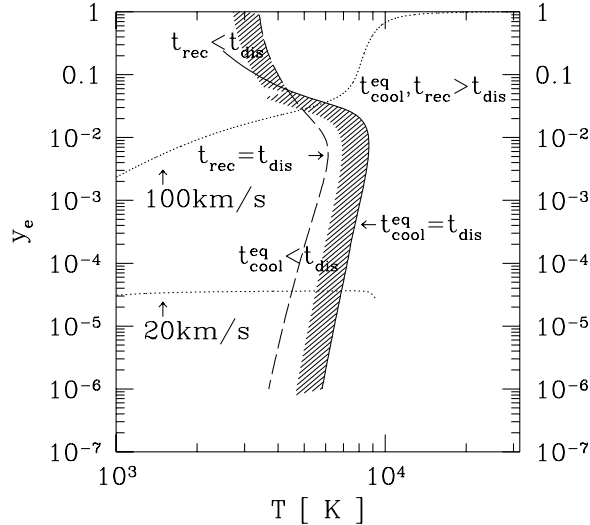}}
   \caption{
        The $y_e$-$T$ plane is divided into two regions, one in  
        which chemical equilibrium is achieved for a fraction of 
        hydrogen molecules and one in which this is not the case. 
    The solid boundary of the shaded region 
    is determined by the condition $t_{\rm cool}= t_{\rm dis}$. 
        The long dashed line denotes 
        the boundary at which $t_{\rm rec}= t_{\rm dis}$ is satisfied. 
        In the shaded region, the chemical equilibrium of 
        hydrogen molecules breaks down, 
        and the fraction $y_{\rm H_2}$ 
                freezes as the temperature and/or ionization degree drops.
        The dotted lines labeled $`100~{\rm km/s}'$
        and $`20~{\rm km/s}'$ 
        denote the evolutionary path of $y_e$.} 
\label{fig2}
\end{figure}

We compare $t_{\rm dis}$ with $t_{\rm rec}$ and $t_{\rm cool}$ in 
the $y_e$-$T$ plane (Fig.~\ref{fig2}). \footnote{Note that Fig.~\ref{fig2} 
is corresponds to Fig.~2 of Susa et al.,\cite{Set} 
but calculated using new reaction rates and cooling function.\cite{GP}
There exist some quantitative diferences, however.  
In particular, the line $t_{\rm rec}= t_{\rm dis}$ changes somewhat. 
This is mainly because 
the dissociative recombination rate of \H2p 
(${\rm H_2}^+ + e \rightarrow 2{\rm H}$) 
is reduced,\cite{GP} and hence the effective \Hbun dissociation rate is 
reduced. 
}
At higher temperature, $t_{\rm dis}$ is shorter than 
$t_{\rm rec}$ and $t_{\rm cool}$, and then  
$y_{\rm H_2} = y_{\rm H_2}^{\rm eq} (y_e, T)$. 
After cooling, in the shaded region, the chemical equilibrium of 
hydrogen molecules breaks down, and $y_{\rm H_2}$ 
almost freezes as the temperature and/or ionization degree drops.
Thus, we can roughly determine $y_{\rm H_2}$ for given $y_e$ and $T$  
in the cooling layers of post-shock flow.  

%%%%%%%%%%%%%%%%%%%%%%%%%%%
% IONIZATION V.S. COOLING
%%%%%%%%%%%%%%%%%%%%%%%%%%%
\subsubsection{Ionization and cooling}

\begin{figure}
   \epsfxsize=8cm
   \centerline{\epsfbox{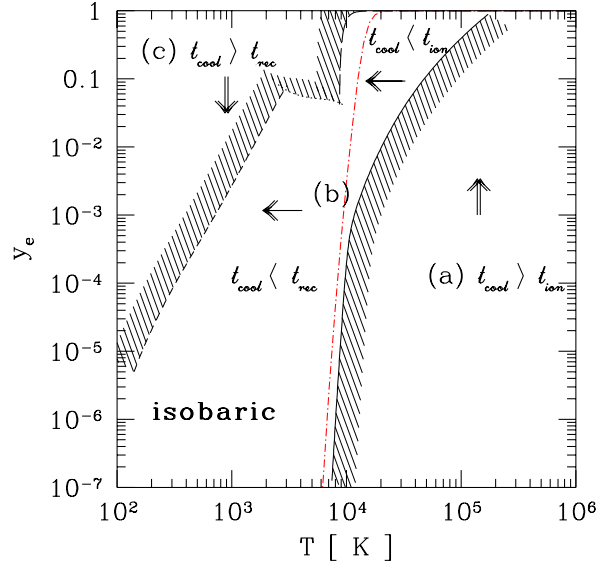}}
   \caption[dummy]{The $y_e$-$T$ plane is divided into several regions
     in which the ratio of the time scales are different from each
     other. The arrows on the plane denote the directions of the 
     evolution of the post-shock layer. 
     The dot-short dashed line denotes the curve 
     on which the equality $t_{\rm rec} = t_{\rm ion}$ holds. 
         The region (a) in which the condition
     $t_{\rm ion} < t_{\rm cool}$ holds is bounded by the solid line. In
     this region, the
     system evolves upward in the $y_e$-$T$ diagram. In
     the region (b), the conditions $t_{\rm cool} < t_{\rm ion}, ~t_{\rm rec}$ 
     always holds. Therefore the temperature drops down
     before the system is further ionized or recombines.
     In the region (c) we have $t_{\rm rec} < t_{\rm cool}$, 
     and the system evolves downward. The region (c) is bounded by 
     the dotted line, short dashed line and the long dashed line.
     These three lines denote the boundary on which the condition
     $t_{\rm cool}=t_{\rm rec}$ holds. The long dashed line is calculated
     by assuming that the cooling rate is dominated by hydrogen line 
     cooling. The dotted line is calculated
     assuming the chemical equilibrium of the hydrogen molecules, 
     and the short dashed line is calculated with $y_{\rm H_2} = 10^{-3}$, 
     which corresponds to the frozen value of the hydrogen molecules 
     for a shock velocity higher than $\sim 40~{\rm km/s}$.} 
\label{fig1}
\end{figure}

Here we investigate the case $t_{\rm ion} < t_{\rm rec}$, 
which corresponds to higher temperature ($T \gsim 10^4$ K), 
and compare the time scale of ionization with that of cooling.
The line along which $t_{\rm ion}/t_{\rm cool}$ is equal to unity is 
indicated by the solid line in Fig.~\ref{fig1}. 
If the initial conditions are given below the line, 
the system is ionized before it cools down, 
because $t_{\rm ion}/t_{\rm cool} < 1$.
Otherwise, the system cools before it is ionized. 
The expected evolutionary path is also expressed by the arrows 
in Fig. \ref{fig1}.
\par
%%%%%% Figure1 %%%%%%%%%%%%
\par
%%%%%

%%%%%%%%%%%%%%%%%%%%%%%%%%%%%%%%%%%%%%%%
% RECOMBINATION V.S. COOLING
%%%%%%%%%%%%%%%%%%%%%%%%%%%%%%%%%%%%%%%%
\subsubsection{Recombination and cooling}

Next we investigate the case $t_{\rm rec} < t_{\rm ion}$, 
which corresponds to lower temperature and examine the balance between 
the recombination process and the cooling process.
The recombination process is important for $T \lsim 10^4~{\rm K}$ .
At $T\lsim 10^4~{\rm K}$, the line cooling of the hydrogen molecules 
dominates the atomic hydrogen line cooling.
Therefore the ratio $t_{\rm rec}/t_{\rm cool}$ depends 
not only on the ionization degree and the temperature 
but also on the fraction of hydrogen molecules. 
In order to draw the line $t_{\rm rec}/t_{\rm cool} = 1$ in the 
$y_e$-$T$ plane, 
we need information on the fraction of hydrogen molecules.
The fraction of ${\rm H_2}$ is determined by the chemical
equilibrium above $\sim 10^4~{\rm K}$. Then, it is approximately frozen 
below $\sim 6\times 10^3~{\rm K}$  through an evolutionary course 
with cooling.
The line along which $t_{\rm rec}/t_{\rm cool} = 1$ is satisfied 
is drawn in Fig.~\ref{fig1}.
In the region above the line, the system recombines before it cools down, and
below the line, the system cools down before the recombination proceeds.
The expected evolutionary path is also indicated in Fig.~\ref{fig1} 
by the arrows.

%%%%%%%%%%%%%%%%%%%%%%%%%%%%%

\subsubsection{Shock diagram}
%%%%%%%%%%%%%%%%%%%%%%%%%
The ratio of any two time scales is independent of 
the total density, and they are determined by the temperature 
and the chemical compositions ($y_e$ and $y_{\rm H_2}$). 
$y_{\rm H_2}$ is almost always determined by $y_e$ and $T$. Therefore,
the ratio of any two time scales just depends on $T$ and $y_e$.
In the following sections, we compare the time scales individually.

The main results are summarized in Fig.~\ref{fig1}. The
evolution of a shock heated system is basically explained 
in the $y_e$-$T$ plane,
which we call ``shock diagram''. The evolutionary path
of the post-shock layer is obtained by tracing the directions of the
arrows on the shock diagram. 
Post-shock gas, which is heated up to high temperature, 
appears in region (a). 
In region (a), the shortest time scale is $t_{\rm ion}$, and 
the system evolves upward and enters region (b). 
In region (b), $t_{\rm cool}$ is the shortest, and 
therefore the system evolves leftward here and comes into region (c). 
In region (c), $t_{\rm rec}$ is the shortest. Then  
the system evolves downward and reenters the region (b), where
it evolves leftward again and crosses the boundary of regions
 (b) and (c).
Finally, the system evolves nearly along the short dashed curve 
which is the boundary of regions (b) and (c).  
For post-shock gas which is not heated above $2\times 10^4~{\rm K}$,
the thermal evolution is somewhat different from the previous case.
In this case, when the system enters region (b) from region (a), 
the ionization degree is so low that the system does not enter 
region (c) at $T \simeq 10^4$~K.  
Then the ionization degree and the fraction of hydrogen molecules do
not ``forget'' the initial conditions of the chemical composition.

\subsubsection{Convergence of $y_e$ and $y_{\rm H_2}$ 
in almost steady postshock flow} 

It is known that in steady post shock flow, both $y_e$ and $y_{\rm H_2}$ 
become the same value at the same temperature. 
This is surprising because ionization equilibrium is not yet achieved 
at all. 
Now, the reason that covergence of $y_e$ and $y_{\rm H_2}$ 
in the steady post-shock flow occurs can be understood using 
Figs.~\ref{fig2} and \ref{fig1}.

At $T \sim 10^4~{\rm K}$, the line 
$t_{\rm rec}=t_{\rm cool}$ becomes nearly vertical, 
since the dominant cooling process is the hydrogen atomic line cooling,
which decreases by many orders below  $T \sim 3 \times 10^4~{\rm K}$. 
If (1) the initial shock velocity is larger than $\sim 40~{\rm km/s}$ or
(2) the initial ionization degree is larger than $\sim 5\times 10^{-2}$
and the initial temperature is larger than $T \sim 10^4~{\rm K}$, then
the evolutionary path should hit this nearly vertical line (the long
dashed line in Fig.~\ref{fig1}). 
Because the line $t_{\rm rec}=t_{\rm cool}$ is nearly vertical at $T
\simeq 10^4~{\rm K}$ and $y_e \gsim 5 \times 10^{-2}$, 
$y_e$ drops to several $\times 10^{-2}$ without changing the temperature, 
and ``forgets'' the initial
conditions, though $y_e$ is not in chemical equilibrium.
The value to which $y_e$ converges ($\sim 5\times 10^{-2}$) is essentially
determined by the equation 
$t_{\rm rec}(y_e,T)=t_{\rm cool}^{\rm eq}(y_e,T)$ at 
$T\simeq 8000~{\rm K}$. 
Here, the temperature $8000$~K is obtained by
the equality $t_{\rm rec}/t_{\rm cool}^{Ly{\alpha}}=1$, where
$t_{\rm cool}^{Ly{\alpha}}$ is the cooling time estimated by the hydrogen
line cooling. 
The equation $t_{\rm rec}/t_{\rm cool}^{Ly{\alpha}}=1$ depends both 
on $y_e$ and on $T$ in general, but it depends only on $T$ for $y_e \ll 1$.

The reason for the convergence of ${\rm H_2}$ also can be understood. 
According to Fig.~\ref{fig2}, ${\rm H_2}$ is still in chemical equilibrium 
just below $T=10^4~{\rm K}$, at which the convergence of $y_e$ takes place.
As a result, the systems that satisfy the previous condition (1) or (2)
experience the same state (same $y_e$, same $y_{\rm H_2}$) 
just below $T=10^4~{\rm K}$. 
Therefore $y_{\rm H_2}$ also converges for different initial conditions.

%%%%%%%%%%%%%%%%%%%%%%%%%%%%%%%%%%%%%%%
% COMPARISON WITH THE NUMERICAL CAL.
%%%%%%%%%%%%%%%%%%%%%%%%%%%%%%%%%%%%%%%
\subsubsection{Comparizon with numerical calculations}

%%%%%%%%%% Figure3 %%%%%%%%%%%%%%
\begin{figure}
   \epsfxsize=8cm
   \centerline{\epsfbox{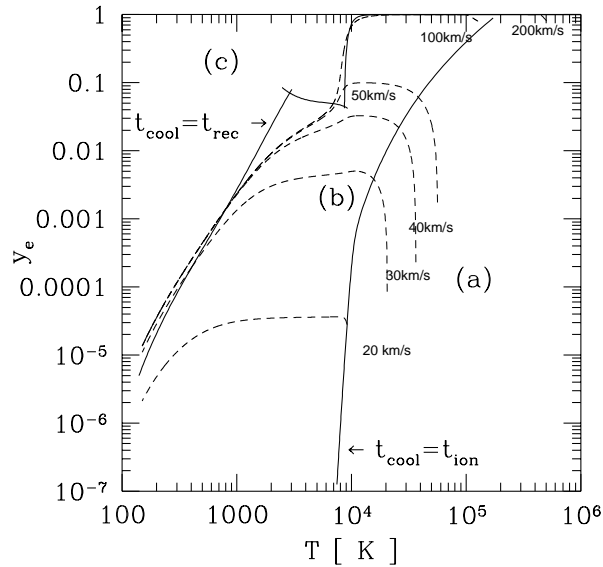}}
   \caption[dummy]{
The evolution of the post-shock layer in the $y_e$-$T$ plane.
The solid lines are the boundaries beyond which the direction of 
evolution is changed, and each line has the same meaning as in 
Fig.~\ref{fig1}. The short dashed lines are the numerical results. 
Each suffix denotes the initial shock velocity.
   }
\label{fig3}
\end{figure}

In this section we compare the results obtained in the previous section
with numerical calculations. 
Figure~$\ref{fig3}$ displays the evolution of $y_e$ as a function of 
the temperature behind the shock front in steady flow.
It is obvious that the expected evolutionary path in Fig.~$\ref{fig1}$
agrees with the numerically calculated path.
We also find that there exists a critical value 
of the initial shock velocity above
which the evolution of $y_e$ below $10^4~{\rm K}$ converges.
The critical value of the shock velocity is $\sim 40~{\rm km/s}$.
This condition for the initial shock velocity is equivalent 
to the condition (1) in the previous subsection.  

Because of the delay of the recombination process compared to 
{\it the line cooling due to the hydrogen molecules}, 
fairly high degree of ionization ($y_e\sim 10^{-1.5}$) is possible 
below $8000 {\rm K}$.
Feeded the electrons, ${\rm H_2}$ also maintains a high abundance  
($y_{\rm H_2}\sim 10^{-2}$)
for $T \lsim {\rm several} \times 10^3~{\rm K}$, then the cloud temperature 
drops to $T\sim 100~{\rm K}$ quickly due to the ${\rm H_2}$ line cooling.
In fact, the predicted evolutionary
paths agree well with the numerically integrated paths for steady
flow.

\subsubsection{Fragmentation of the post-shock layer}

\begin{figure}
   \epsfxsize=8cm
   \centerline{\epsfbox{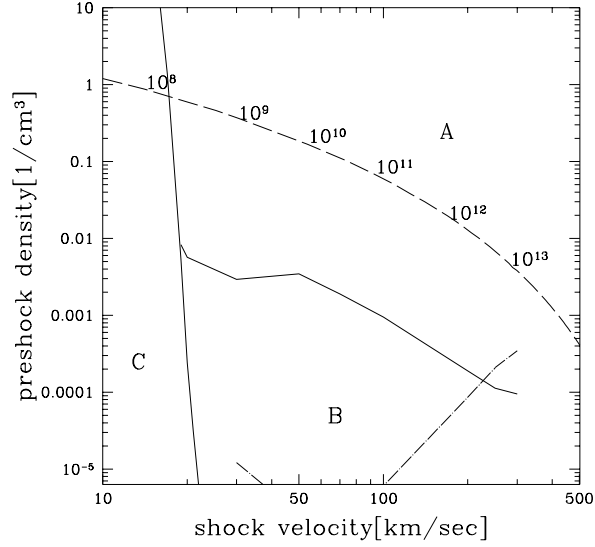}}
   \caption{
Estimated fragmentation epoch. 
Pre-shock density-shock velocity plane, which is divided into
  three regions. Region 
  $A$ is where fragmentation occurs at low temperature $\sim$ 100~K,
  and region $B$ is where it occurs at several thousand kelvin. 
  In region $C$
  the post-shock temperature cannot rise above several thousand kelvin. 
  Typical values of the shock velocity and the pre-shock density at the
  collapse in the standard CDM model are overplotted (dashed line). 
  Numbers plotted above this line indicate the mass scales in unit of
  solar mass.
  The dot-dashed line divides the plane according to
  whether fragmentation occurs within 1 Gyr. 
  The time elapsed since the point at which the shock front cross 
  until fragmentation is calculated with $z=5$.
  Most of the galactic sized
  clouds are cooled within 1 Gyr. Adapted from Yamada and
  Nishi (1998). }
\label{masako}
\end{figure}

To estimate the fragmentation condition of the post-shock layer, 
we can consider $t_{\rm cool}$ as $t_{\rm dyn}$ and 
$t_{\rm ff}$ as $t_{\rm frag}$, where $t_{\rm ff}$ is the free-fall time 
scale, which coresponds to the growth time scale of gravitational 
instability.\cite{EE}
With the standard CDM cosmological model, 
we can estimate the fragmentation epoch. \cite{YN,UN} 
As shown in Fig.~\ref{masako},\cite{YN} primordial gas clouds, which collapse 
in the early universe, fragment after they are clooded down to 
about 100~K if $M \gsim 10^8 M_\odot$. 
In the case that $M \lsim 10^8 M_\odot$, the post-shock temperature is lower 
than $10^4$~K, and \Hbun is formed only slightly. 
So, from the initial state in which $t_{\rm cool} > t_{\rm ff}$, 
the cloud evolves by gravitational force without cooling. 
Then the cloud will be virialized and cannot fragment, because 
it cannot be thin enough for fragmentation.  
%%%%%%%%%%%%%%%%%%%%

\section{Collapse and fagmentaion of filamentary clouds} 

A cooled sheet  fragments more easily  into 
filaments\cite{MNH,U} 
than into spherical clouds.
As the virial temperature of a filament is essentially 
{\it constant}, its evolution is much 
different from that of  a spherical cloud, whose virial temperature 
increases as the cloud contracts.
In this section, we investigate the evolution of filamentary 
clouds of primordial gas and estimate the fragmentation epoch. 

\subsection{Basic equations}

We investigate the dynamical and thermal evolution with a 
1-zone approximation for the simplicity.\cite{Uet} 
Then, basic equations are as follows. 
For an isothermal cylinder there is a critical line density 
$M_c(T)$ given  by
\begin{equation}
M_c(T)= \frac{2k_{B}T}{\mu m_{\rm H} G} ,            \label{k2}     
\end{equation}
where $ T, \mu $, and $m_{\rm H}$ represent the isothermal
 temperature, the mean molecular weight  and  the mass of
 hydrogen atoms, respectively.\cite{Ost} 
An isothermal cylinder with line density $M$ larger than
 $M_c(T)$  has no equilibrium  and collapses, while  
an isothermal cylinder with $M$ less  
than $M_c(T)$  expands  if the external pressure does not exist.

When the  filamentary cloud is formed as a result of  the fragmentation
 of the sheet, the gas pressure is comparable to 
the gravitational force. Hence the effect of the gas pressure must be 
included in the equation of motion of the cylinder like cloud. 
To know the dynamical evolution of such a cylinder we begin with the virial 
equation for a cylinder, 
\begin{eqnarray}
\frac{1}{2} \frac{d^2 I}{d t^2}=2 T+2 {\it \Pi} - G M^2 , \label{v1}
\end{eqnarray}
where
\begin{eqnarray}
  I = \int_V \rho r^2 dV   , \;\;\;
  T = \int_V \frac{1}{2} \rho u^2 dV  ,\;\;\;
{\it \Pi} = \int_V p dV   , 
\end{eqnarray}
and the integration is effected over the volume per unit length of 
 the cylinder, $V$.
Here we assume that the cylinder is uniform and has constant 
density $\rho$, scale radius $R=\sqrt{M/ \pi \rho}$, and  temperature $T$.  
Then $I$,~$T$ and $\Pi$ introduced above become
\begin{eqnarray}
  I = \frac{1}{2}M R^2  ,\;\;\;     
  T = \frac{1}{4}M \left ( \frac{dR}{dt} \right )^2  ,\;\;\;
{\it \Pi} = \frac{k_B T}{\mu m_{\rm H}} M \; .  \label{P2}  
\end{eqnarray}
Substituting Eq.~(\ref{P2}) in Eq.~(\ref{v1}) we obtain
\begin{eqnarray}
\frac{d^2 R}{dt^2}&=&-\frac{2 G}{R} ( M-M_c(T) ) \; .  \label{eq2}  
\end{eqnarray}
We describe the evolution of the cylinder by this equation.

The time evolution of the cloud temperature is described by the energy 
equation given by
\begin{equation}
\frac{d \varepsilon}{dt}=-P \frac{d}{dt}\frac{1}{\rho}-\Lambda_{\rm rad} 
- \Lambda_{\rm chem}  \;,   \label{z2}
\end{equation} 
where $\varepsilon$, $\Lambda_{\rm rad}$ and $\Lambda_{\rm chem}$ are 
the thermal energy per unit mass, the cooling by the radiation, and
the cooling/heating by chemical reactions, respectively.

Assuming that the collapsing cylinder fragments 
when the condition (\ref{s1}) is satisfied, we estimate 
the epoch of fragmentatioin and the mass of fragments.  
The estimated mass scales correspond to the stellar mass scale 
($1 M_\odot \sim 100 M_\odot)$. 

\subsection{Analytical estimate of the minimum fragment mass}  

Now, we estimate the minimum fragment mass analytically. 
As the cylinder collapses isothermally as a function of time 
in the later stage,  
we can set $d \varepsilon/dt \sim 0$ in Eq.~(\ref{z2}) and
from this we obtain $t_{\rm dyn} \sim (\gamma-1) \: t_{\rm cool}$, where
$\gamma=7/5$ for a diatomic molecular gas.  
Since the cloud cools by the line emissions of hydrogen molecules and
is optically thick to these line emissions,
$t_{\rm cool}$ can be estimated by 
\begin{equation}
t_{\rm cool} \sim  \frac{ \frac{1}{\gamma-1} \frac{M}{\mu m_{\rm H}} k_B T}
              {2\pi R \sigma T^4 \frac{\Delta \nu}{\nu} \alpha_c } ,
\label{A1}
\end{equation}
where $\sigma=2 \pi^5 k_B^4 /15 h^3 c^2 $ is the Stefan-Boltzmann constant, 
and $\alpha_c$ is the effective number of line emissions.   
Since the optical depth is $\sim 100$ at most, the wing of the line profile
may not be important,\cite{RadiPro} and the line
broadening is determined by Doppler broadening as
${\Delta \nu}/{\nu}={v_{\rm H_2} }/{c}
=\sqrt{ {k_B T}/{m_{\rm H} c^2} } $.
The temperature  of the cylinder is estimated by the virial
temperature as
\begin{equation}
k_B T=\frac{1}{2}\mu m_{\rm H} G M  . \label{A2}
\end{equation}
From the above equations we can estimate the fragment mass as  
\begin{eqnarray}
M_{\rm frag}  \sim  2\pi R M    
 \sim   \sqrt{ \frac{1}{\alpha_c} }
\frac{1}{\mu^{9/4} }\frac{m_{Pl}^3}{m_{\rm H}^2}   ,  \label{A3}
\end{eqnarray}
where $m_{Pl}=\sqrt{hc/G}$ is the Planck mass.
Equation (\ref{A3}) implies that  $M_{\rm frag}$ is essentially equal to
the Chandrasekhar mass ($\sim m_{Pl}^3/m_p^2$, where $m_p$ is proton mass).  
\par
To this point we have not considered other heating processes, such as
 shocks and turbulence, which  likely occur in a collapsing cloud.
All these processes tend to halt the collapse, so that
the epoch of fragmentation  becomes  earlier, and, as a result, the mass of
fragments increases.  
Thus Eq.~(\ref{A3}) gives a lower limit on the mass of the fragment
which is formed from a primordial gas cloud. 
Our estimate shows that even a primordial gas cloud which
cools most effectively by hydrogen molecules cannot fragment into 
a smaller mass than the expressed in Eq.~(\ref{A3}).

For smaller line densities the gas pressure force halts the collapse  
before three-body reactions 
 {\it completely} convert atomic hydrogen into molecules.\cite{PSS}
In these cases, Eq.~(\ref{A3}) is not applicable, because 
clouds are optically thin to line emissions if the 
molecular fraction is smaller than about $10^{-1}$, and
chemical heating by ${\rm H_2}$ formation must be considered when 
hydrogen molecules are formed  by three-body reactions.
These clouds fragment at lower density, 
in other words at larger scale radius. 
Since  $M_{\rm frag}$ is proportional to the scale radius at fragmentation,
this leads to the larger fragment mass.

In reality, numerical simulations suggest that 
primordial cylindrical clouds fragment at lower density, and 
the estimated masses of fragmets are larger  
than the Chandrasekhar mass.\cite{U,NM} 

\section{Evolution of cloud cores}

Now we study the evolution process of cloud cores. 
After they collapse into stellar cores, true stars can form. 

\subsection{Quasi-static contraction}

The manner in which primordial clouds of stellar mass, which are fragments of
filamentary clouds, evolve after the fragmentation depends on their
efficiency of cooling, i.e., the ratio of the free-fall time scale
$t_{\rm{ff}}$ to the cooling time scale $t_{\rm{cool}}$.   
If $t_{\rm{ff}}<t_{\rm{cool}}$ (i.e., cooling is not effective), while
the cloud cools it has enough time to adjust itself to a new
hydrostatic configuration.  
Then it contracts, maintaining a nearly hydrostatic equilibrium (i.e. 
Kelvin-Helmholtz contraction). 
On the other hand, in the case $t_{\rm{ff}}>t_{\rm{cool}}$, the cloud
collapses dynamically in its free-fall time scale after substantial
cooling, and no hydrostatic equilibrium is established. 

At the time of the fragmentation of filamentary clouds, the two time scales of
fragments, $t_{\rm{cool}}$ and $t_{\rm{ff}}$, are of the same order of
magnitude.\cite{Uet} 
Thus we investigate quasi-static contraction of cloud cores 
as a first step. \cite{Oet}

\subsubsection{Radiative processes in primordial molecular cloud cores}

We consider spherically symmetric gas clouds with stellar mass ($\sim 1
M_{\odot}$) (`molecular cores') composed only of H$_2$ molecules, because
the hydrogen atoms are converted into molecular form before the
fragmentation for these low mass fragments.\cite{Uet}   
We can neglect helium, since it is thermally inert at these low
temperatures.  
As mentioned above, these clouds lose their thermal energy by radiative
cooling via the H$_2$ rotovibrational lines, and presumably contract
into protostars.  
 
In this section, we describe our calculation scheme for the luminosity and
cooling rate of a spherically symmetric cloud. 
The specific intensity $I_{\nu}~({\rm ergs}~{\rm sec^{-1}}~{\rm
cm^{-2}}~{\rm sr^{-1}}$\break${\rm Hz^{-1}})$ along a ray is calculated by
solving the radiative transfer equation,\cite{RadiPro} 
\begin{equation}
\frac{dI_{\nu}}{ds}=-\alpha_{\nu}I_{\nu}+j_{\nu} .
\label{eq:tr}
\end{equation}
Here $s$ is the displacement along the ray, and $\alpha_{\nu}$ and  
$j_{\nu}$ are the absorption and emission coefficients, respectively. 
These coefficients can be written using Einstein $A$- and
$B$-coefficients:
\begin{eqnarray}
j_{\nu}&=&\frac{h \nu}{4 \pi} n_{2} A_{21} \phi (\nu),\\
\alpha_{\nu}&=&\frac{h \nu}{4 \pi}\phi (\nu) (n_{1}B_{12}-n_{2}B_{21}).
\end{eqnarray}
Here $n_{1}$ and $n_{2}$ are the number densities of the molecules in
the lower and upper levels of the transition, respectively, and
$\phi(\nu)$ is the line profile function.  
In our calculation, the line center optical depth of the lines across a
cloud $\tau_{\rm{c}}$ is no more than about several hundred. 
Since the Lorentz wings of a line become more important than its Doppler
core only when $\tau_{\rm{c}}$ is as large as $10^{3}$, we only consider
line broadening owing to the thermal Doppler effect.

Note that the critical number density $n_{\rm{cr}}$, where radiative and
collisional deexicitation rates become equal for molecular hydrogen, is
$ \sim 10^{4}~{\rm cm^{-3}}$.  
Thus, for a typical number density of fragments $(\sim 10^{10}~{\rm cm^{-3}}
\gg n_{\rm{cr}})$, almost all excited molecules would be deexcited by
collision with other molecules. 
Therefore, collisional local thermodynamic equilibrium (LTE) is
established, and the scattering of photons can be neglected. 
These properties make the transfer problem of H$_2$ lines fairly tractable. 

\subsubsection{Evolution of fragments} 
As initial conditions, we assume polytropic gas spheres in hydrostatic
equilibrium; namely the density/temperature distribution is represented
by the Emden function of polytropic index $N$.
Clouds are cut off at the radius $r_{s}$, where the density
falls off by a factor of $10^{-3}$ from the central value. 
Therefore, the parameters characterizing initial conditions are the effective
polytropic index $N$, the total mass of the cloud $M$, and the number
density $n_{h}$ at the half mass radius. 
For a given initial configuration, we can obtain the specific entropy
distribution $s(m,t=0)$.   

We calculate quasi-static contraction as follows. 
First we calculate the cooling rate $\Lambda(m)$, 
and advance the specific entropy distribution $s(m,t)$
using the heat equation  
\begin{equation}
\frac{\partial s}{\partial t}=-\frac{1}{T} \Lambda(m).
\end{equation}
With the new entropy distribution, we find the new equilibrium
configuration using the equations of hydrostatic equilibrium.  

For given states of clouds, we compare two time scales, the
free-fall time
\begin{equation}
t_{\rm{ff}}(m)\equiv \sqrt{\frac{3\pi}{32 G \bar{\rho}(m)}},
\end{equation}
and the time scale of quasi-static contraction,  
\begin{equation}
t_{\rm{qsc}}(m)\equiv \rho /\left(\frac{\partial \rho}{\partial t}
\right)_{m,{\rm quasi \hbox{-}static}}.
\end{equation}
Here, $\bar{\rho}(m)$ is the mean density at the mass coordinate
$m$, and
each quantity is evaluated at the coordinate $m$. From the virial theorem,
we know that $t_{\rm{qsc}}$ is of the same order of magnitude as 
$t_{\rm{cool}}$.

\begin{figure}
  \epsfxsize= 8.8cm
  \centerline{\epsfbox{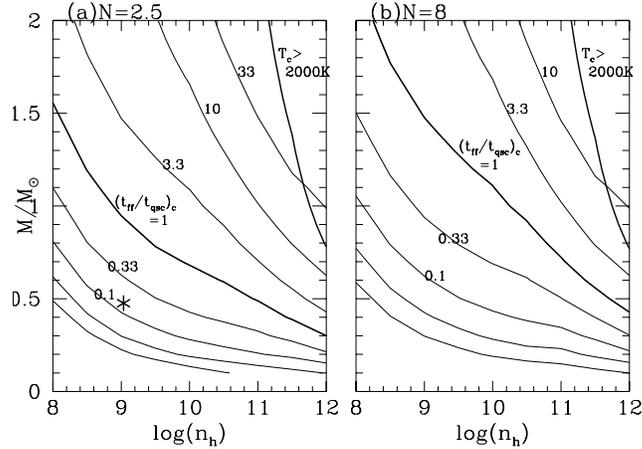}}
\caption{Contours of the ratio $t_{\rm{ff}}/t_{\rm{qsc}}$ at the centers of
  clouds of polytropic index (a) $N=2.5$ and (b) $N=8$. 
  The contour spacing is logarithmic with an increment of 0.5. 
  The upper-right regions correspond to temperatures $T>2000~{\rm
  K}$, where hydrogen molecules begin to dissociate. 
  Adapted from Omukai et al. (1998)}
\label{fo1}
\end{figure}

Figure~\ref{fo1} displays the contours  of 
the ratio $t_{\rm{ff}}/t_{\rm{qsc}}$ at the center ($m=0$) of clouds with
(a) $N=2.5$ (adiabatic stratification) and (b) $N=8$. 
%These $M-n_{h}$ planes include more general states than those of fragments. 
Regions around the curves $t_{\rm{ff}}/t_{\rm{qsc}}=1$ are expected to be 
initial states of fragments. 
Although Fig.~\ref{fo1} displays the values at the
center, this ratio does not change significantly within a cloud, except
near the surface. 
The upper-right regions correspond to higher central temperature $>
2000~{\rm K}$, where the molecular hydrogen begins to dissociate. 
From the two panels of Fig.~\ref{fo1}, we can see that the ratio
$t_{\rm{ff}}/t_{\rm{qsc}}$ depends only slightly on the effective polytropic
indices $N$, rather than on masses and densities. 
Note that for clouds of the same mass, the ratio
$t_{\rm{ff}}/t_{\rm{qsc}}$ is greater for the denser cloud. 
This implies that as a cloud contracts, the ratio $t_{\rm{ff}}/t_{\rm{qsc}}$
becomes larger. 
In particular, for fragments of filamentary gas clouds, for which
$t_{\rm{ff}} \sim t_{\rm{qsc}}$ initially, $t_{\rm{qsc}}$ becomes
shorter than $t_{\rm{ff}}$ as they contract; i.e., such clouds collapse
dynamically. 
Note that when $t_{\rm{qsc}}<t_{\rm{ff}}$, clouds do not contract 
quasi-statically, and $t_{\rm{qsc}}$ does not possess the meaning of the
collapse time scale. 
In this case, $t_{\rm{qsc}}$ merely measures the time scale of cooling.  

\begin{figure}
  \epsfxsize= 8.8cm
  \centerline{\epsfbox{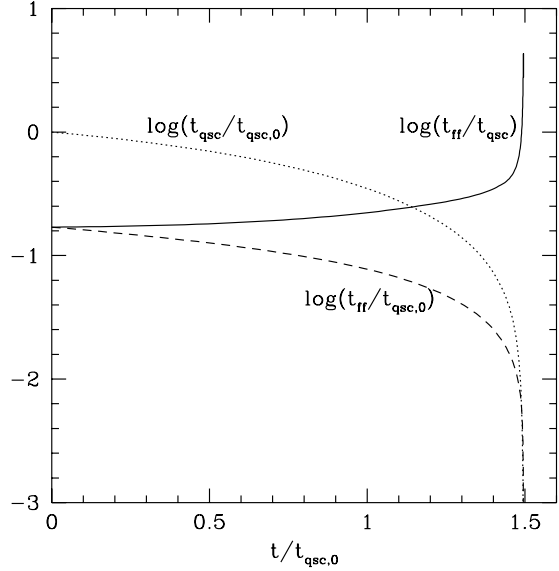}}
\caption{Evolution of the free-fall time $t_{\rm{ff}}$ and dynamical time for
  quasi-static contraction $t_{\rm{qsc}}$ and their ratio 
  at the center for a cloud which contracts quasi-statically at the beginning
  (asterisk of Fig. 5). The time scales $t_{\rm{ff}}$ and
  $t_{\rm{qsc}}$ and the time ellapsed from the begining $t$ are all
  normalized by $t_{\rm{qsc},0}$, the initial value of $t_{\rm{qsc}}$.
  This figure indeed shows that the cloud begins to
  collapse dynamically after $\sim 1.5 ~t_{\rm{qsc},0}$. 
  Adapted from Omukai et al. (1998)}
\label{fo2}
\end{figure}

Figure~\ref{fo2} exhibits the change of time scales 
and their ratio for the quasi-static contraction of a cloud 
which is initially at the asterisk position of Fig.~\ref{fo1}. 
We can see that the cloud indeed collapses dynamically after it contracts 
quasi-statically to some extent.

\subsubsection{Line profile and cooling rate}

We have shown that the stellar mass molecular cores 
collapse dynamically, although some lines are optically thick at the line
center. 
This fact indicates that the efficiency of cooling by H$_{2}$ lines does
not fall significantly even in such a situation, in contrast to the case 
of continuous radiation.   
The reason for this can be interpreted as follows.

\begin{figure}
  \epsfxsize= 8.8cm
  \centerline{\epsfbox{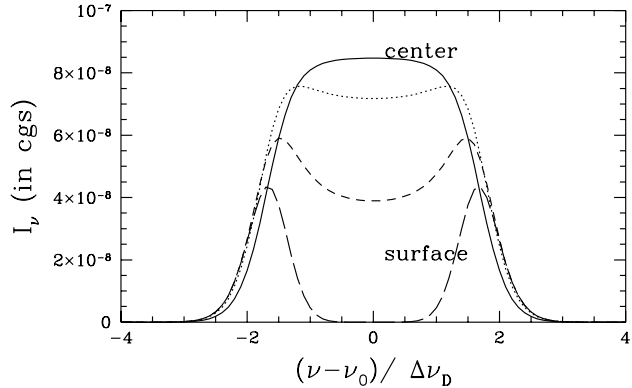}}
\caption{Processed line profiles of an optically thick line along a
  ray passing through the center of a cloud at 
  radii containing 0\% (center, solid line), 1.4\% (dotted line),
9.9\% (short dashed line), and 100\% (surface, long dashed line) of
  the total mass. 
The optical depth at the line-center frequency $\tau_{\rm{c}}=38$, 46,
  54 and 76, respectively,
  measured from the surface of incidence. 
 The line corresponds to the transition from $(v=1,~j=3)$ to
  $(v=0,~j=1)$. 
  The parameters of the cloud are the same as in Fig.~2. 
  The abscissa represents the frequency from the line center 
  normalized by the Doppler width at the center of the cloud $\Delta
  \nu_{\rm{D,c}}$. Adapted form Omukai et al. (1998)}
\label{fo3}
\end{figure}

Figure~\ref{fo3} displays the change of 
a processed profile of a single optically
thick line (optical depth at line center across the cloud is 76) along a ray
passing through the center of the cloud. 
Since the cloud is optically thick at line center, the specific
intensity at line center is saturated with the blackbody value
determined from the local temperature.  
On the other hand, the optical depth at the wing is smaller than unity. 
The final processed profile at the cloud surface is 
double peaked, and the peak frequencies correspond to 
frequencies where the optical depth across the core $\tau_{\nu,\rm{core}}>1$
(i.e., the radiation is saturated with the value of the core
temperature) and that of envelope $\tau_{\nu,\rm{env}}<1$ (i.e., the
radiation can be larger than the value of the envelope temperature). 
The height (i.e., the normalization of intensity) and width of these
peaks are of the same order of magnitude as the blackbody value
determined from the temperature near the center, not around the surface
of the cloud.  
The reduction factors from the central values are smaller than about 10. 
Owing to the effect described above, which is a distinctive feature of
line cooling, the temperature around the cloud center 
can be `seen' from outside the cloud, even when lines are optically
thick at line center and the surface temperature is relatively
low. Accordingly, its cooling proceeds  
efficiently, as long as there are enough frequency ranges 
where $\tau_{\nu,\rm{core}}>1$ and $\tau_{\nu,\rm{env}}<1$.
Therefore, the decrease in the efficiency of cooling 
caused by the low surface temperature is not severe in the line
cooling case, in contrast to the case of cooling by continuum.

\subsection{Dynamical collapse of primordial protostellar clouds}

As discussed above, although the cooling time of a cloud is longer than 
the its free-fall time initially, the cooling time at the center 
becomes shorter than the free-fall time, and the cloud begins 
to dynamical collapse after some quasi-static contraction.  
In this section we investigate the dynamical collapse of primordial 
protostellar clouds and the formation of stellar cores.\cite{Oet} 

\subsubsection{Method of calculation}

We assume spherical symmetry for the simplicity. 
Then the Lagrangian equations describing dynamical collapse are 
as follows. 

\noindent
Equation of continuity: 
\be
{\partial m \over \partial r} = 4 \pi r^2 \rho, 
\ee
equatioin of motion: 
\be
{D v \over D t} = - 4 \pi r^2 {\partial p \over \partial m} 
- {G m \over r^2}, 
\ee
equation of energy: 
\be
{D \epsilon \over D t} = -p {D \over D t} 
\left( {1\over \rho} \right) -{\Lambda \over \rho}, 
\ee
equation of state: 
\be
p = (\gamma_{ad} - 1) \rho \epsilon.
\ee
In the above expressions, $m$, $\rho$, $v$, $p$, $\epsilon$, $\Lambda$ 
and $\gamma_{ad}$ 
are the mass within a radius $r$, the density, velocity, pressure, 
thermal energy per unit mass, net cooling rate per unit volume 
and the adiabatic coefficient, respectively. 

The net cooling rate $\Lambda$ consists of two parts, 
the radiative part $\Lambda_{\rm rad}$ and the chemical part 
$\Lambda_{\rm chem}$. $\Lambda_{\rm rad}$ can be written as  
\be
\Lambda_{\rm rad}(m) = \rho {\partial L(m) \over \partial m}, 
\ee
where $L(m)$ is the luminosity and is obtained by solving radiative 
transfer equations. 
We consider \Hbun lines and continuum components which consist of 
\Hbun collision-induced absorption (CIA), H$^-$ bound-free absorption, 
Lyman continuum absorption, etc. as sources of opacity. 
$\Lambda_{\rm chem}$ is given by 
\be
\Lambda_{\rm chem} (m) = \rho {\partial \epsilon_{\rm chem} (m) 
\over \partial t}, 
\ee
where $\epsilon_{\rm chem}$ is the chemical binding energy per unit
mass. 

\subsubsection{Dynamical evolution of protostellar clouds} 

\begin{figure}
  \epsfysize= 12cm
  \centerline{\epsfbox{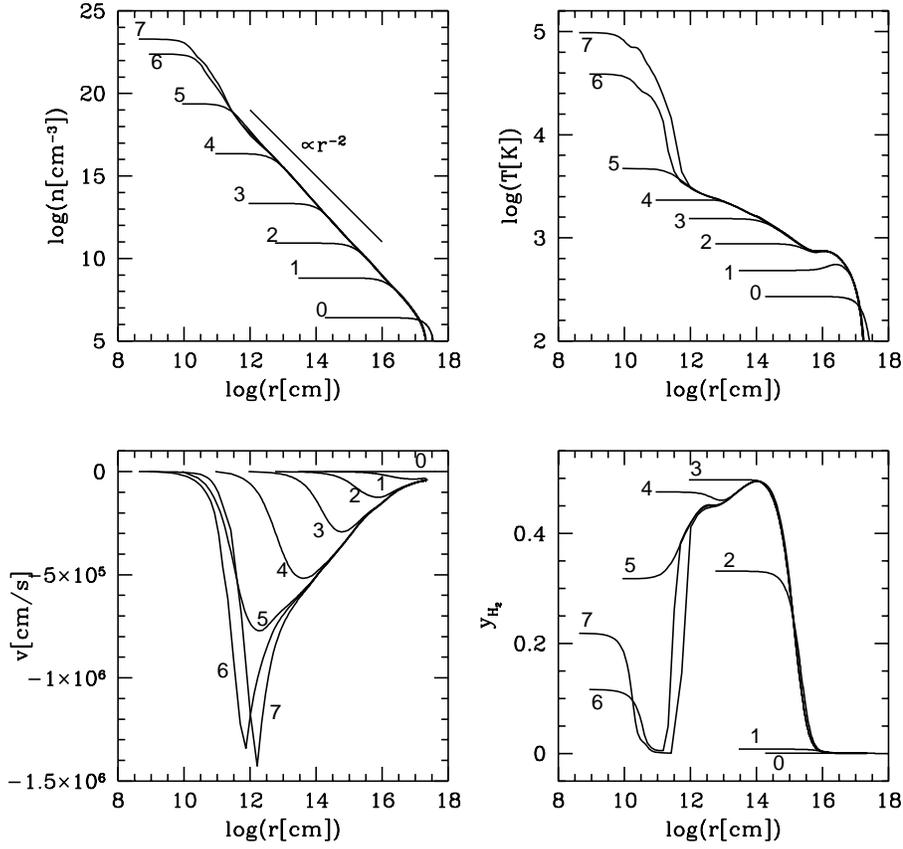}}
\caption{
The radial profiles of the physical quantities. (a) The number density 
($n \equiv \rho/m_{\rm H}$, where $m_{\rm H}$ is the mass of the
hydrogen atom), (b) temperature, (c) velocity, and (d) the H$_2$ 
concentration are given as functions of the radial distance. 
The numbers in the figure denote the evolutionary sequences. 
`0' corresponds to the initial time, and `1' to 5.7 $\times~10^5$~yrs 
after starting calculation. Here, 
three-body processes are active in the central region and temperature 
inversion occurs because of efficient \Hbun line cooling. 
`2'corresponds to 8.5 $\times~10^3$ yrs after state 1: 
The cloud becomes optically 
thick to some \Hbun lines. `3' corresponds to 2.8 $\times~10^2$~yrs 
after state 2: 
The central region becomes fully molecular. `4' corresponds to 12~yrs 
after state 3: 
The central region become optically thick to the \Hbun CIA continuum.
`5' corresponds to 0.32~yrs after state 4: 
\Hbun dissociation becomes efficient. 
`6' corresponds to 2.4 $\times~10^{-2}$~yrs after state 5: 
Shortly after the core 
formation. `7' corresponds to 4.1 $\times~10^{-2}$~yrs after state 6. 
Adapted from Omukai and Nishi (1998)}
\label{fcore}
\end{figure}

The evolutionary sequences of the radial profiles of number density, 
temperature, velocity 
and \Hbun concentration distributions are illustrated in Fig. \ref{fcore}. 

The collapse proceeds almost self-similaly and is analogous to 
the Larson-Penston similarity solution\cite{LP} until the 
central region reashes the stellar density. 
The dashed line in the figure indicates the slope of density gradient 
of $-2$. The slope in this case is slightly steeper and its value is 
aboud $-2.2$, which correspond to the effective adiabatic index is 1.1. 

At higher density, almost all hydrogens are converted to \Hbun by 
efficient three-body processe. 
But the total mass of central molecular part is about 1 $M_\odot$ which 
hardly depends on total cloud mass. 
When the central number density reaches about 
$3 \times 10^{13}~{\rm cm}^{-3}$ and  
 the central temperature reaches about 1600 K, the hydrogen molecules 
begin to dissociate gradually. 
But at almost the same time, \Hbun CIA continuum cooling becomes 
efficient. This continuum cooling is so strong to stop and 
even to reverse dissociation. 
After a small central part of the cloud becomes opaque to 
CIA continuum (3 $\times 10^{16} ~{\rm cm}^{-3}$, 2000 K), 
the effective dissocication begins. 

\begin{figure}
  \epsfysize= 8.8cm
  \centerline{\epsfbox{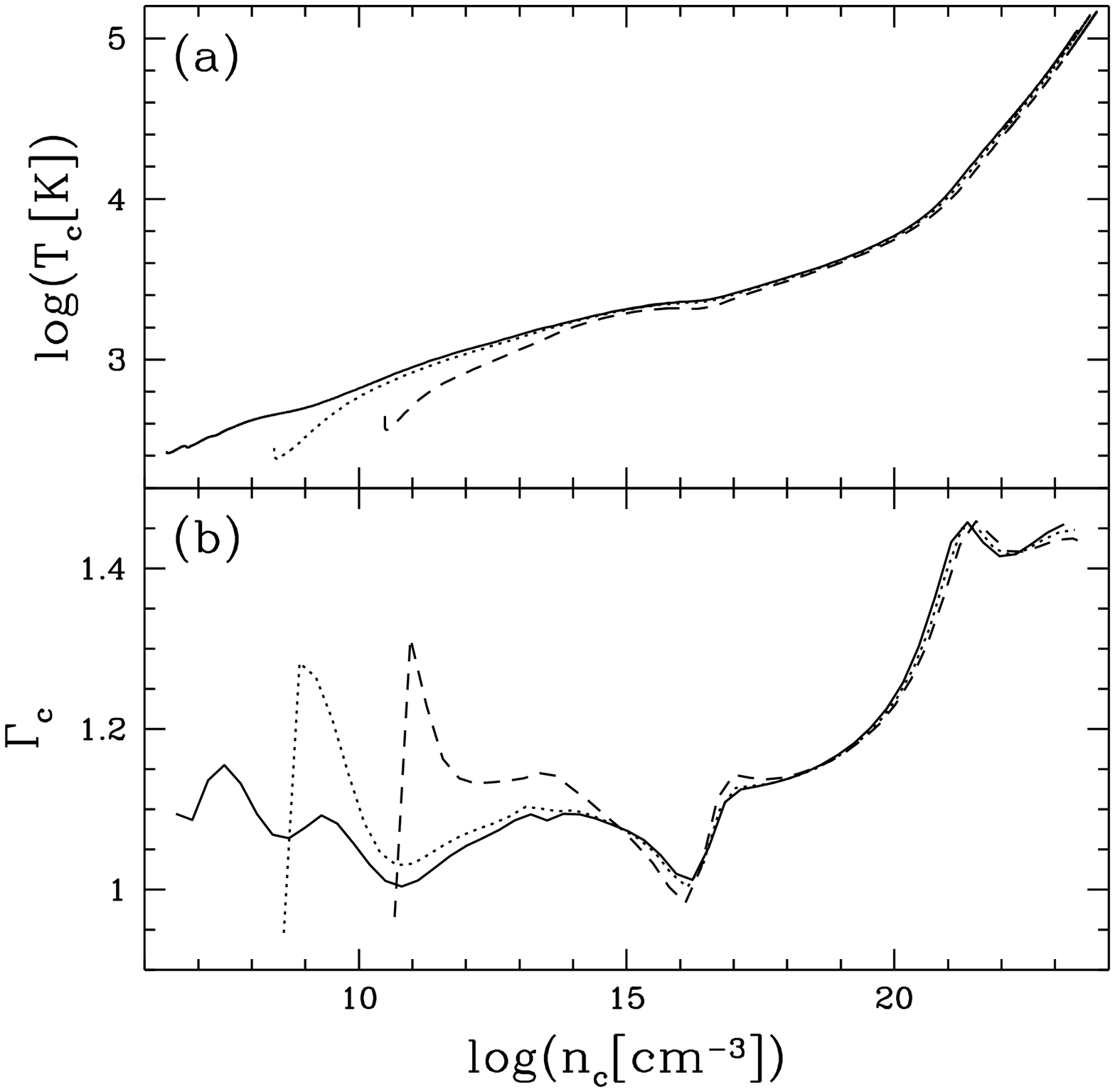}}
\caption{
The temperature and effective `adiabatic index'  
$\Gamma$ at the central regions are plotted 
against the central number density $n_c~[{\rm cm}^{-3}]$ 
for clouds with various masses. 
(a) The central temperature $T_c$. (b) The effective `adiabatic index' 
$\Gamma_c$ in the central regions. 
In both panels, the total masses are $10^2 M_\odot$ (solid lines), 
$10 M_\odot$ (dotted lines) and $1 M_\odot$ (dashued lines). 
Adapted from Omukai and Nishi (1998)}
\label{fcen}
\end{figure}

The evolution of the temperature and the effective `adiabatic index' 
${\Gamma = {\partial \ln p / \partial t \over 
\partial \ln \rho / \partial t}}$ at the center for various total mass 
cases are illustrated in Fig. \ref{fcen}. 
We can see the central value $\Gamma$ is 
about 1.1 almost throughout the collapse. 
After some contraction, evolution paths of the central regions are 
hardly affected by the total mass. 

When most of the \Hbun molecules are dissociated, $\Gamma$ rises above 
the critical value 4/3. 
After the central part of the cloud contracts almost adiabatically 
to some extent, a hydrostatic core with very small mass 
($\sim 5 \times 10^{-3} M_\odot$) forms at the center. 
At that time, the central number density and temperature are  
$\sim 10^{22}~{\rm cm}^{-3}$ and $\sim 3 \times 10^4$~K. 
These values are almost the same as those of present-day 
stellar cores at the formation epoch. 

But before the formation of the stellar core, no transit core is formed, 
in contrast to the case of present-day star formation. 
And the estimated mass accreation rate in the case of  primordial star 
formation is much different from that of present-day star formation 
($\sim 10^{-5} M_\odot ~{\rm yr}^{-1}$), as shown below. 
Befor stellar core formation, the radial profiles of the density are 
well fitted by the Larson-Penston-type similarity solution 
with $\gamma = 1.09$. Assuming that the evolution of an envelope after 
stellar core formation is described by the same similarity
solution, we can estimate the evolution of the core mass. 
With the similarity solution after the central density becoming infinite, 
we have for the central stellar mass\cite{YSS} 
\be
M_* = 0.11 M_\odot \left( {t \over 1~{\rm yr}} \right),  
\ee
and we can estimate the mass accretion rate as 
\ba
\dot M_* &=& 8.3 \times 10^{-2} M_\odot~{\rm yr}^{-1} 
\left( {t \over 1~{\rm yr}} \right)^{-0.27}, \\ 
&=& 3.7 \times 10^{-2} M_\odot~{\rm yr}^{-1} 
\left( {M_* \over M_\odot} \right)^{-0.37}. 
\ea

The mass accretion rate is very large and diminishes with time. 
The main reason for this large mass accretion rate is that  
the temperature of primordial gas clouds ($\sim 10^3$ K) is higher 
(which is the result of weak cooling) than that of the present-day 
molecular clouds ($\sim 10$ K). 
The final stellar mass may be large because of the large mass 
accretion rate. 

\section{Summary and discussion}

We investigated the thermal and the dynamical 
evolution of primordial gas clouds  
in the universe after decoupling and discussed the 
fragmentation process. 
It is necessary to study the evolution of the clouds 
considering the change of cloud shapes.  
Comparing the time scale of the dynamical evolution 
with that of fragmentation,  
we estimated the typical fragmentation scale. 

\subsection{A formation scenario of first luminous objects}
We propose a formation scenario for first luminous objects 
(Fig. \ref{sc}):

\begin{figure}
\epsfysize=15cm % fix the y-dimension and scales x-dim. to y-dim.
\hspace{0.7cm}\epsfbox{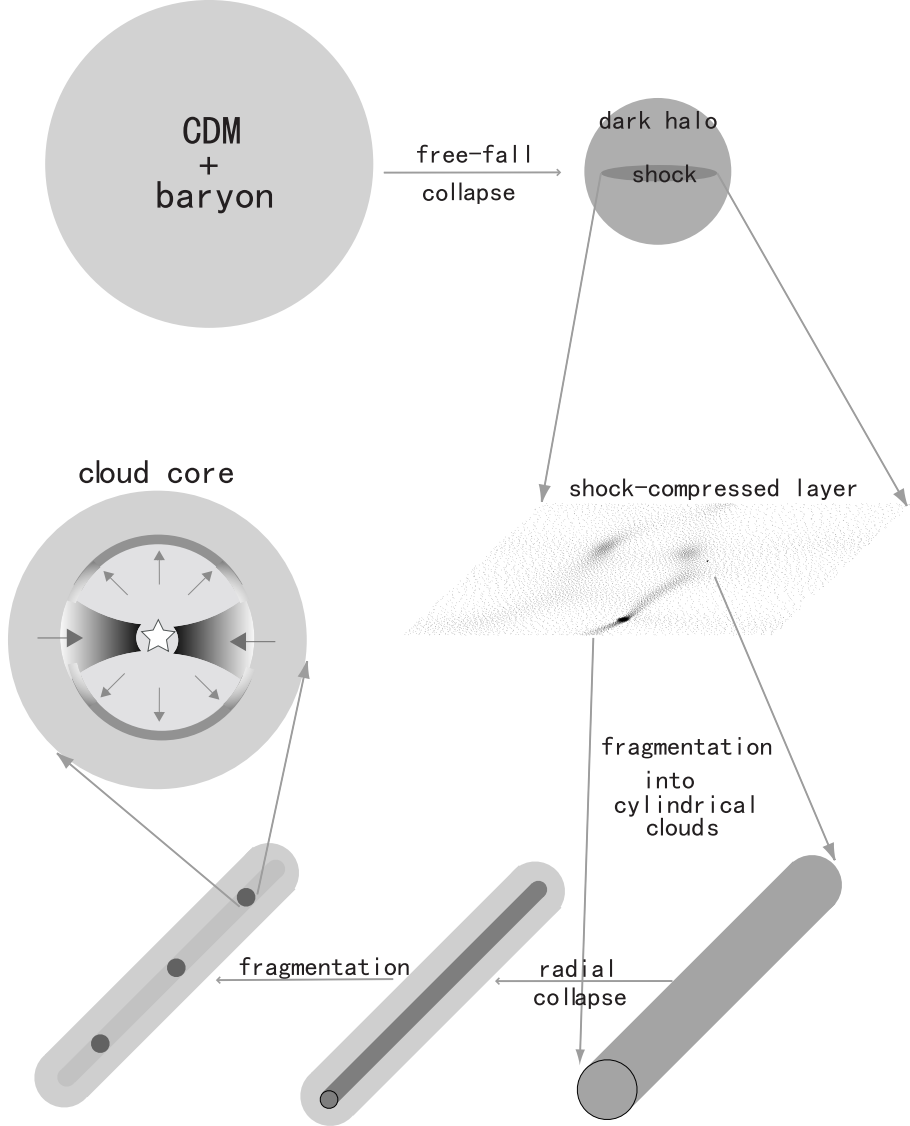}
\caption[h] {The formation scenario for first luminous objects. 
\label{sc}}
\end{figure}

\begin{list}{}{}
\item[(1)] By pancake collapse of the over-density regions 
or collision between clouds in potential wells, 
quasi-plane shocks form. 

\item[(2)] If the shock-heated temperature is higher than about $10^4$~K, 
the post-shock gas cools by ${\rm H_2}$ line cooling, 
and the shock-compressed layer fragments into cylindrical clouds 
when $t_{\rm dyn} \sim t_{\rm frag}$ ($T$ is several hundred K). 

\item[(3)] The cylindrical cloud collapses dynamically once more 
and fragments into cloud cores when $t_{\rm dyn} \sim t_{\rm frag}$. 

\item[(4)] A primordial star forms in a cloud core. 

\item[(5)] In the case that the shock-heated temperature is higher than 
about $10^4$~K, 
many fragments can be formed almost simultaneously. 
Thus, many massive stars may be formed 
within the time scale of the life of massive stars. 

\item[(6)] Luminous objects are formed. 
\end{list}

The condition (5) is achieved when a massive cloud ($\gsim 10^8 M_\odot$) 
collapses or subgalactic clouds collide with each other in the potential 
of massive objects, which has the virial velocity larger than about 
40 km/s.  
In any case, luminous objects can be formed after 
massive objects ($M \gsim 10^8 M_\odot$) collapse.  

In the case that the shock-heated temperature is lower than about $10^4$~K, 
ionization can not occur and \Hbun molecules are formed only 
with cosmological relic electrons. Hence $y_{\rm H_2}$ is small and 
cooling by \Hbun is weak, 
and the cloud cannot fragment into many small clouds. 
In this case, one or a few massive stars may be formed 
around the cloud center. 
After the first generation of massive stars are formed, 
the cooling of the remaining part of the cloud becomes 
inefficient, since hydrogen molecules are dissociated by UV photons 
radiated by massive stars. Thus, the star formation rate is strongly 
self-regulated.\cite{ON2} Moreover, the binding energy of such 
a small mass cloud 
is small, and the cloud is blown off or disturbed severely by 
a super nova explosion, and as a result, the formation of the 
next generation of stars is greatly delayed. 
Then it is unlikely for isolated small mass clouds to become 
luminous objects. 

\section*{Acknowledgements}
We thank H. Sato and T. Nakamura continuous encouragement.
This work is supported in part by Research Fellowships of the Japan Society 
for the Promotion of Science for Young Scientists, No. 2370 (HS), 
6894 (HU) and Grant-in-Aid 
for Scientific Research on Priority Areas (No. 10147105) (RN) and 
Grant-in-Aid for Scientific Research from the Ministry of Education,
Science, Sports, and Culture, No. 08740170 (RN).

\def\apj{Astrophys. J.}
\def\aap{Astron. Astrophys.}
\def\mnras{Mon. Not R. Astron. Soc.}

\end{document}